\documentclass[journal, draftcls, one column, 12pt]{IEEEtran}
\usepackage{color}
\usepackage{leftidx}
\usepackage{cite}
\usepackage{multirow}
\usepackage{subfigure}
\usepackage{graphicx,amssymb,amstext,amsmath}
\usepackage{cleveref}
\usepackage[ruled,vlined,lined,ruled,linesnumbered]{algorithm2e}
\usepackage{url}
\usepackage[belowskip=-5pt,aboveskip=0pt]{caption}
\begin{document}
	
	\title{Joint 3D Deployment and Resource Allocation for UAV-assisted Integrated Communication and Localization}
	
	\author{Suzhi~Bi, Jiaxing~Yu, Zheyuan~Yang, Xiaohui~Lin, and Yuan Wu%,~\IEEEmembership{Senior Member,~IEEE}
		\thanks{
			\IEEEcompsocthanksitem
			S. Bi, J. Yu and X. Lin are with the State Key Laboratory of Radio Frequency Heterogeneous Integration,  Shenzhen University, Shenzhen, Guangdong, China, 51800 (e-mail:\{bsz,~xhlin\}@szu.edu.cn, yuyxx9@163.com). S. Bi is also with the Department of Broadband Communication, Peng Cheng Laboratory, Shenzhen, China, 518066. Z. Yang is with the Department of Information Engineering, The Chinese University of Hong Kong, Hong Kong SAR (e-mail: yz019@ie.cuhk.edu.hk). Y. Wu is with the State Key Laboratory of Internet of Things for Smart City and the Department of Computer and Information Science, University of Macau, Macau, SAR, China (e-mail: yuanwu@um.edu.mo). }
		}
	
	%\IEEEpeerreviewmaketitle
	\maketitle

	\begin{abstract}
		In this paper, we investigate an unmanned aerial vehicle (UAV)-assisted integrated communication and localization network in emergency scenarios where a single UAV is deployed as both an airborne base station (BS) and  anchor node to assist ground BSs in communication and localization services. We formulate an optimization problem  to maximize the sum  communication rate of all users under localization accuracy constraints by jointly optimizing the 3D position of the UAV, and communication bandwidth and power allocation of the UAV and ground BSs. To address the intractable localization accuracy constraints, we introduce a new performance metric  and geometrically characterize the UAV feasible deployment  region in which the  localization  accuracy constraints are satisfied. Accordingly, we  combine  Gibbs sampling (GS) and block coordinate descent (BCD) techniques to tackle the non-convex joint optimization problem.  Numerical results show that the proposed method attains almost identical rate performance as the meta-heuristic benchmark method while reducing the CPU time by $89.3\%$.
	\end{abstract}
	\begin{IEEEkeywords}
		UAV, integrated communication and localization, network deployment, resource allocation
	\end{IEEEkeywords}
	
	\captionsetup[figure]{labelfont={bf},name={Fig.},labelsep=period}
	\par
	
	\captionsetup[figure]{labelfont={bf},name={Fig.},labelsep=period}
	\par

\newpage

\section{Introduction}
5G/B5G technologies provide not only fast communication connections but also accurate localization service to mobile users when localization based on global navigation satellite system is unreliable, enabling many smart applications such as autonomous driving. However, in emergent scenarios such as post-disaster rescue,  cellular communication and localization service can be severely degraded due to hardware malfunctions or severe blockage and scatterings on the ground. A cost-effective solution is deploying unmanned aerial vehicles (UAVs) to establish reliable line-of-sight (LoS) air-to-ground links.

UAV-based aerial base stations (BSs) can assist the ground network in improving both communication and localization services. There have been many studies on deploying UAVs to improve communication coverage \cite{ref3}, data rate \cite{ref4}, communication delay and energy consumption performance \cite{ref17,ref19}, etc. A common design methodology is to jointly optimize the flight trajectory (or deployment locations) and resource allocation (e.g., power, bandwidth, and transmission time) of the UAVs to increase the communication link capacity under a limited onboard battery. For UAV-assisted localizations, the problem becomes more complicated. In 4G/5G standards, the time difference of arrival (TDoA) based method is widely used for high-precision localization \cite{ref7}. With TDoA localization method, determining a unique 3D localization estimate requires at least four collaborative positioning anchor nodes.  Besides, the localization accuracy depends on the deployment and resource allocation of all the participating anchor nodes \cite{ref7}. Given the aerial anchor node positions, \cite{ref8} investigates the resource allocation in a UAV-assisted vehicular network, where the authors derive the Camer-Rao Lower Bound (CRLB) of localization accuracy and propose an optimal resource allocation method to minimize the CRLB using semi-positive definite programming. Optimizing the UAV anchor position, however, is much more challenging because the CRLB is non-convex and analytically intractable in the anchor position. Existing works mostly use meta-heuristic methods for optimizing the UAV anchor location. For example, \cite{ref9} uses a particle swarm optimization method to design the UAV trajectory for localizing vehicles. \cite{ref11} considers a UAV-assisted vehicle positioning network and proposes an iterative method based on Taylor expansion. \cite{ref12} optimizes the UAV trajectory using a genetic algorithm to reduce energy consumption on data collection from ground sensors under positioning accuracy constraints. Overall, the existing methods fail to analytically characterize UAV anchor placement solutions, and thus suffer from high computational delay in emergency network deployment.

On the other hand, existing methods mostly consider using a UAV for either providing communication or localization service. In practice, this leads to high hardware cost (i.e., using more UAVs) and large delay when deploying emergency networks. Using one UAV for providing both services is cost-effective, however, also challenging due to the potential conflicts in deciding the optimal deployment and the resource sharing between communication and localization services.

In this paper, we consider reusing one UAV to collaborate with three ground BSs for providing both communication and localization services to ground users. In particular, we maximize the sum communication rate of all users under localization accuracy constraints by jointly optimizing the 3D position of the UAV and network resource allocation. Instead of CRLB, we propose a variant of the \emph{D-optimality criterion} as the metric of localization accuracy, and derive a \emph{closed-form expression} of the feasible 3D UAV placement region. With the concise characterization of localization constraint, we propose an efficient algorithm based on Gibbs sampling to solve the non-convex joint optimization problem. Simulation results show that the proposed method achieves almost identical rate performance as the meta-heuristic benchmark method while reducing the computation delay by $89.3\%$.

\section{System Model and Problem Formulation}
In this paper, we consider a network where one UAV and three BSs collaboratively provide downlink communication and localization services to $K$ ground users. The coordinates of the $n$th BS, the $k$th user, and the UAV are denoted as $\mathbf{b}_{n}=[x_{n}^{b},y_{n}^{b},h_{n}^{b}],n\in\left\{ 1,2,3\right\} $, $\mathbf{w}_k=[x_{k},y_{k},h_{k}],k\in  \left \{1,2,\cdots,K\right\}$, $\mathbf{u}=[x_u,y_u,h_u]$, respectively.\footnote{Accurate user location is unknown and to be estimated, where $\mathbf{w}_k$ is an assumed user location, e.g., from initial coarse location estimation\cite{ref7}.} Suppose that the BSs and UAV communicate with ground users using Orthogonal Frequency Division Multiplexing (OFDM) signals. Here, we consider a unified Rician fading model to account for the dissimilar blockage, shadowing, and scattering effects of the ground-to-ground (G2G) and air-to-ground (A2G) channels, specified by the different continuous power ratio of the multipath fading components $\omega_G, \omega_A \in [0,1]$. In particular, we consider that the G2G channel has a weak LoS link (i.e., $\omega_{G} \approx 1$) and the A2G channel has a dominant LoS link (i.e., $\omega_A$ close to zero) because of the high altitude of the UAV transmitter.

Let $|h_{nk}|^2$ denote the channel power gain between the $n$th BS and the $k$th user with $\mathbb{E}\left[|h_{nk}|^2\right] = \frac{\beta}{||\mathbf{b}_{n}-\mathbf{w}_{k}||^\iota} \triangleq g_{nk}$, where $\beta$ is the reference path loss at 1 meter and $\iota$ is the path loss exponent of the G2G channel. When $\omega_G =0$, i.e., pure LoS channel, $|h_{nk}|^2 = g_{nk}$ is a deterministic value. Otherwise, for $\omega_G \in (0,1]$, let $Z \triangleq \frac{|h_{nk}|^2}{g_{nk} \omega_G/2}$, we can show that $Z$ follows a non-central $\chi^2$ distribution with the noncentrality parameter $\lambda_G = \frac{2\left(1-\omega_G\right)}{\omega_G}$ and degree of freedom equals $2$, denoted as $Z \sim F\left(z;2,\lambda_G\right)$ with $F\left(\cdot\right)$ being the cumulative distribution function (CDF). Consider an outage probability tolerance $\varepsilon$, the achievable rate of the G2G channel is \cite{ref16}
	\begin{equation}
		R_{nk}=s_{nk}B\cdot\log_{2}\left(1+\frac{F^{-1}\left(\varepsilon;2,\lambda_G\right) \beta P_{nk}}{s_{nk}B N_0 ||\mathbf{b}_{n}-\mathbf{w}_{k}||^{\iota}} \right). \label{rateG}\\			
	\end{equation}	
Here, $B$ is the communication bandwidth and $s_{nk}\in [0,1]$ is the fraction of bandwidth allocated by the $n$th BS to the $k$th user, $F^{-1}(\cdot)$ is the inverse function of the CDF, $P_{nk}$ is the transmit power, and $N_0$ is the noise power density. Similarly, the achievable outage communication rate of A2G channel is
	\begin{equation}	
		R_{uk}=s_{uk}B\cdot\log_{2}\left(1+\frac{F^{-1}\left(\varepsilon;2,\lambda_A\right)\beta P_{uk}}{s_{uk}B N_0||\mathbf{u}-\mathbf{w}_{k}||^{\tilde{\iota}}} \right),\label{rateA}
	\end{equation}
	where $\lambda_A=\frac{2\left(1-\omega_A\right)}{\omega_A}$, $s_{uk}$ is the fraction of bandwidth allocated by UAV to the $k$th user, and $\tilde{\iota}$ is the pathloss exponent.
	
	In the UAV-assisted localization network, a user computes its position using the localization signals broadcast by the BSs and UAV. We set BS 1 as the reference localization anchor. For a tagged user k, the TDoA vector can be expressed as
	\begin{equation}
		\begin{aligned}
			&\mathbf{\Delta\tau}=[\Delta\tau_{2k},\Delta\tau_{3k},\Delta\tau_{uk}].
		\end{aligned}
	\end{equation}
	Here, $\Delta\tau_{ik}=(d_{ik}-d_{1k})/c+\delta_{i}-\delta_{1},i\in\{2,3,u\}$, $d_{ik}$ is the distance between anchor $i$ and user $k$, $c = 3\cdot 10^8$ m/s is the speed of light, $\delta_{i} \sim \mathcal{N}(0,\sigma_i^2)$ is the  measurement error of time of arrival (ToA) of the signal from anchor $i$. Then, the covariance matrix of the TDoA vector is \cite{ref12}
	\begin{align}
		C=
		\begin{bmatrix}
			\sigma_{1}^2+\sigma_{2}^2&\sigma_{1}^2&\sigma_{1}^2\\
			\sigma_{1}^2&\sigma_{1}^2+\sigma_{3}^2&\sigma_{1}^2\\
			\sigma_{1}^2&\sigma_{1}^2&\sigma_{1}^2+\sigma_{u}^2
		\end{bmatrix},
	\end{align}
	where $\sigma_{n}^2=\psi\cdot \frac{\bar{B}N_0 ||\mathbf{b}_{n}-\mathbf{w}_{k}||^{\iota}}{\beta \bar{P_{n}} }+\sigma^{2}_{nlos} \triangleq \frac{\psi}{SNR_n} + \sigma^{2}_{nlos}, n\in \{1,2,3\}$ is the ToA measurement variance from the $n$th BS anchor, $\psi$ is a constant related to the positioning signal design parameters \cite{ref12}, $\bar{B}$ is the bandwidth of positioning signal,  $\bar{P}_n$ is the positioning power of the $n$th BS, $\sigma^{2}_{nlos}$ is the variance of ToA measurement error resulting from the multi-path non-LoS effect. Meanwhile, $\sigma_u^2 =\psi\cdot \frac{\bar{B} N_0 ||\mathbf{u}-\mathbf{w}_{k}||^{\tilde{\iota}}}{\beta \bar{P_{u}}} \triangleq \frac{\psi}{SNR_u}$ is the variance of ToA measurement from the UAV anchor, $\bar{P}_u$ is the positioning power of the UAV.
	
	For simplicity of exposition, we define unit vectors $\mathbf{q}_n=\frac{\mathbf{b}_n-\mathbf{w}_k}{||\mathbf{b}_n-\mathbf{w}_k||}=[q_{n1},q_{n2},q_{n3}]$, $\mathbf{q}_u=\frac{\mathbf{u}-\mathbf{w}_k}{||\mathbf{u}-\mathbf{w}_k||}=[q_{u1},q_{u2},q_{u3}]$. Then, the Jacobian matrix of the TDoA vector is \cite{chen}
	\begin{equation}
		H=
		\begin{bmatrix}
			\frac{\partial \Delta\tau_{21} }{\partial x_{k}}&\frac{\partial  \Delta\tau_{21} }{\partial y_{k}}&\frac{\partial  \Delta\tau_{21} }{\partial h_{k}}\\
			\frac{\partial  \Delta\tau_{31} }{\partial x_{k}}&\frac{\partial  \Delta\tau_{31} }{\partial y_{k}}&\frac{\partial  \Delta\tau_{31} }{\partial h_{k}}\\
			\frac{\partial  \Delta\tau_{u1} }{\partial x_{k}}&\frac{\partial  \Delta\tau_{u1} }{\partial y_{k}}&\frac{\partial  \Delta\tau_{u1} }{\partial h_{k}}
		\end{bmatrix}=
		\begin{bmatrix}
			\mathbf{q}_2-\mathbf{q}_1\\
			\mathbf{q}_3-\mathbf{q}_1\\
			\mathbf{q}_u-\mathbf{q}_1
		\end{bmatrix}.	
	\end{equation}
	We can calculate the Fisher information matrix (FIM) of the user position by $F= H^{T}C^{-1}H$. In this paper, instead of CRLB, we use D-optimality criterion \cite{ref20} as the metric of positioning accuracy to derive a tractable deployment solution
	\begin{equation}
		opt\text{-}D=det(F)=det(H^{T}C^{-1}H)=det^{2}(H)/det(C),
	\end{equation}
	where $det(\cdot)$ represents the determinant of matrix.

	We aim to maximize the sum data rate of users by jointly optimizing the UAV's position $\mathbf{u}$, communication power  ${P}=\{P_{11},P_{12},\cdots,P_{uN}\}$, positioning power ${\bar{P}}=\{\bar{P}_1,\bar{P}_2,\bar{P}_3,\bar{P}_u\}$ and communication bandwidth allocation ${S}=\{s_{11},s_{12},\cdots,s_{uN}\}$ while ensuring the minimum required localization accuracy and communication rate of each user:
	\begin{subequations}
		\begin{align}
			\mathrm{(P1):}
			\quad &  \underset{ {P},{S},\mathbf{u},{\bar{P}}}{\text{max}} \ \
			\sum_{k=1}^{K} R_{k}  \nonumber \\
			s.t.
			\quad & \sum_{k=1}^{K}P_{jk}+\bar{P_{j}} = P_{max},\forall j \in\{1,2,3,u\}, \label{con:power} \\
			\quad & \sum_{k=1}^{K} s_{jk}=1,\forall j \in\{1,2,3,u\}, \label{con:bandwidth}\\
			\quad & R_k\geq R_{th}, \forall k\in\{1,2,\cdots,K\},  \label{con:rate}\\
			\quad & opt\text{-}D(\mathbf{u},{\bar{P}},\mathbf{w_k})\geq \epsilon_k, \forall k\in\{1,2,\cdots,K\},  \label{con:accuracy}\\
          	\quad & P_{jk}, \bar{P_{j}}, s_{jk} \geq 0, \forall j,k, \label{con:fea}
		\end{align}	
	\end{subequations}
	where $R_k=\sum_{j=1}^{3}R_{jk}+R_{uk}$ is the data rate of user $k$, (\ref{con:power}) denotes the transmit power constraint of the BSs and UAV, (\ref{con:bandwidth}) is the bandwidth allocation constraint, (\ref{con:rate}) denotes the minimum communication rate requirement $R_{th}$ of all users, and (\ref{con:accuracy}) denotes the positioning accuracy requirement with threshold $\epsilon_k$ for user $k$. (P1) is difficult to solve because of the non-convex objective function and constraints (\ref{con:rate}) and (\ref{con:accuracy}).

	\section{Characterization of Positioning Constraint }
\begin{figure}[t]
		\centering
		\includegraphics[scale=0.7]{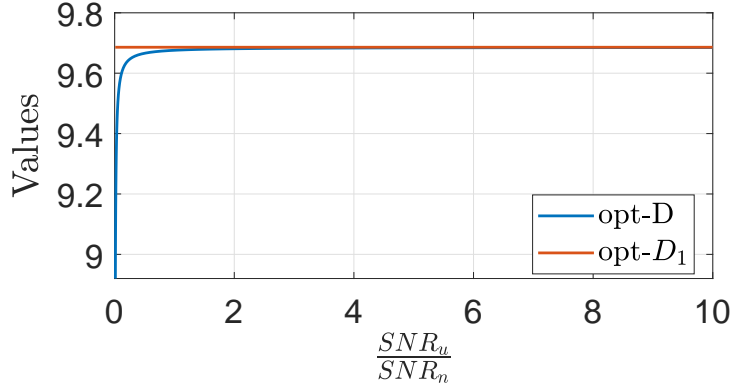}
		\caption{Values of $opt\text{-}D$ and $opt\text{-}D_1$ under different $\frac{SNR_u}{SNR_{n}}$ under the setup in the Simulation section. The $SNR_n$'s are assumed equal.}
		\label{fig2}
	\end{figure}

	In this section, we simplify the positioning accuracy constraint (\ref{con:accuracy}). Firstly,  we can express $det(C)$ as
	\begin{equation}
		\begin{aligned}
			det(C)=\sigma_{1}^{2}\sigma_{2}^{2}\sigma_{3}^{2}+\sigma_{u}^{2}(\sigma_{2}^{2}\sigma_{3}^{2}+\sigma_{1}^{2}\sigma_{3}^{2}+\sigma_{1}^{2}\sigma_{2}^{2})\triangleq D_{1}+D_{2}
		\end{aligned}.
	\end{equation}
    Suppose that $SNR_u \gg SNR_n$, $\forall n$, or the multipath G2G channel owns large measurement variance $\sigma^2_{nlos}$, we have $\sigma_n^2 \gg \sigma_u^2$, $\forall n$,  such that $D_1 \gg D_2$ and the D-optimality metric can be accurately approximated by the following $opt\text{-}D_1$ metric
	\begin{equation}
		opt\text{-}D_{1}=det^{2}(H)/D_{1}=det^{2}(H)/\sigma_{1}^{2}\sigma_{2}^{2}\sigma_{3}^{2}.
	\end{equation}	
In Fig.~\ref{fig2}, we plot the values of $opt\text{-}D_{1}$ and $opt\text{-}D$ under varying $\frac{SNR_u}{SNR_{n}}$, where we find that $opt\text{-}D_{1} \approx opt\text{-}D$ when $\frac{SNR_u}{SNR_{n}}$ is greater than $0.5$. Even with $\frac{SNR_u}{SNR_{n}}=0.1$, the gap between $opt\text{-}D_{1}$ and $opt\text{-}D$ is less than $2\%$. For analytical tractability, we use $opt\text{-}D_1$ to replace $opt\text{-}D$ in (\ref{con:accuracy}), where $opt\text{-}D_1$ is unrelated to UAV's positioning power. Constraint (\ref{con:accuracy}) then turns into
	\begin{equation}
		det^{2}(H)/D_{1}\geq \epsilon_k,\forall k \label{con:positioningconstrain}.
	\end{equation}

	\textbf{Case 1}: When $det(H)>0$, (\ref{con:positioningconstrain}) is equivalent to
	\begin{equation}
		\alpha_{1}(x_{u}-x_{k})+\alpha_{2}(y_{u}-y_{k})+\alpha_{3}(h_{u}-h_{k})\geq \tilde{\epsilon}_{1}||\mathbf{u}-\mathbf{w}_{k}||, \label{con:temp2}
	\end{equation}
	where $\tilde{\epsilon}_{1}=\sqrt{\epsilon_k D_{1}}+\alpha_{1}q_{11}+\alpha_{2}q_{12}+\alpha_{3}q_{13}$ and
	\begin{equation}
		\begin{aligned}
			\alpha_{1}=(q_{22}-q_{12})(q_{33}-q_{13})-(q_{23}-q_{13})(q_{32}-q_{12}),\\
			\alpha_{2}=(q_{23}-q_{13})(q_{31}-q_{11})-(q_{21}-q_{11})(q_{33}-q_{13}),\\
			\alpha_{3}=(q_{21}-q_{11})(q_{32}-q_{12})-(q_{22}-q_{12})(q_{31}-q_{11}).
		\end{aligned}
	\end{equation}
	(\ref{con:temp2}) is a second-order cone with the user location being the vertex when $\tilde{\epsilon}_{1}>0$. Note that (\ref{con:temp2}) can be expressed as
	\begin{equation}
		\alpha_{1}q_{u1}+\alpha_{2}q_{u2}+\alpha_{3}q_{u3}\geq \tilde{\epsilon}_{1},
	\end{equation}
	which can be viewed as the intersection of the half-space $P= \left \{ (x,y,h)|\alpha_{1}x+\alpha_{2}y+\alpha_{3}h\geq \tilde{\epsilon}_{1} \right \} $ and sphere $O=\left \{ (x,y,h)|x^2+y^2+h^2=1 \right \}$. To have an intersection, the distance from the origin to the half-space should satisfy $\frac{\tilde{\epsilon}_{1}}{\sqrt{\alpha_{1}^2+\alpha_{2}^2+\alpha_{3}^2}} \leq 1$. Therefore, the accuracy $\epsilon_k$ satisfies:
	\begin{equation}
		\sqrt{\epsilon_k D_1} \leq c_1-c_2 \label{22},
	\end{equation}
	where $c_1=\sqrt{\alpha_{1}^2+\alpha_{2}^2+\alpha_{3}^2}$, $c_2=\alpha_{1}q_{11}+\alpha_{2}q_{12}+\alpha_{3}q_{13}$.
	
	From (\ref{con:temp2}), the 2D feasible region of UAV with a fixed altitude $h_u$ is within an ellipse defined by:
	\begin{equation}
		\begin{aligned}
			&( \tilde{\epsilon}_{1}^2-\alpha_{1}^2)(x_u-x_k)^2-2\alpha_{1}\alpha_{2}(x_u-x_k)(y_u-y_k)+\\
			&( \tilde{\epsilon}_{1}^2-\alpha_{2}^2)(y_u-y_k)^2-2\alpha_{1}\alpha_{3}(x_u-x_k)(h_u-h_k)-\\
			&2\alpha_{2}\alpha_{3}(y_u-y_k)(h_u-h_k)+( \tilde{\epsilon}_{1}^2-\alpha_{3}^2)(h_u-h_k)^2=0. \label{equ:feasible}		
		\end{aligned}
	\end{equation}
	For a  quadratic curve $Q=\{(x,y)|ax^2+bxy+cy^2+dx+ey+f=0\}$, $Q$ is an ellipse  if and only if  the eccentricity $b^2-4ac<0.$
	Comparing with (\ref{equ:feasible}), the condition is $(2\alpha_{1}\alpha_{2})^2-4( \tilde{\epsilon}_{1}^2-\alpha_{1}^2)( \tilde{\epsilon}_{1}^2-\alpha_{2}^2)<0$, which is independent to UAV's flying altitude $h_k$. Accordingly, a sufficient condition for the feasible region in (\ref{equ:feasible}) to be an ellipse is $\sqrt{\alpha_{1}^2+\alpha_{2}^2}< \tilde{\epsilon}_{1}$, or equivalently $\sqrt{\tilde{\epsilon}_{1}D_1}> c_3 - c_2$, where $c_3 \triangleq \sqrt{\alpha_{1}^2+\alpha_{2}^2}$. Combined with the result in (\ref{22}), when the accuracy parameter $\epsilon_k$ satisfies
	\begin{equation}
		c_{3}-c_{2} < \sqrt{\epsilon_k  D_{1}} < c_1-c_2, \label{con:fea1}
	\end{equation}
  the positioning constraint (\ref{con:positioningconstrain}) is feasible and the 3D feasible region of UAV is a cone specified in (\ref{con:temp2}).

	\textbf{Case 2}: When $det(H)<0$, (\ref{con:positioningconstrain}) can be written as
	\begin{equation}
		\alpha_{1}(x_{u}-x_{k})+\alpha_{2}(y_{u}-y_{k})+\alpha_{3}(h_{u}-h_{k})\leq  \tilde{\epsilon}_{2}||\mathbf{u}-\mathbf{w}_{k}||, \label{con:temp3}
	\end{equation}
	where $\tilde{\epsilon}_{2}=-\sqrt{\epsilon_k D_{1}}+ c_2$. When $\tilde{\epsilon}_{1}<0$, (\ref{con:temp3}) specifies a second-order cone with the user location being the vertex. Following the similar analysis in Case 1, when $\epsilon_k$ satisfies
	\begin{equation}
		c_{3} + c_{2} < \sqrt{\epsilon_k  D_{1}} < c_1 + c_2, \label{con:fea2}
	\end{equation}
  the positioning constraint (\ref{con:positioningconstrain}) is feasible and the 3D feasible region of UAV is a cone specified in (\ref{con:temp3}).

  Notice that given the positions of the BSs, $det(H)$ is only related to the position of the UAV $\mathbf{u}$. Here, we neglect the case with $det(H)= 0$ (i.e., the user and all the positioning anchors are coplanar) because the UAV has evidently higher altitude than the three ground anchors in practice. Within a connected domain $\mathcal{U}$ where $det(H)\neq 0$ holds for all $\mathbf{u} \in \mathcal{U}$, $det(H)$ is either everywhere positive or everywhere negative for all $\mathbf{u} \in \mathcal{U}$. This leads to a practical method to determine the sign of $det(H)$. That is, for a user $k$ located at $\mathbf{w}_k = \left[x_k,y_k,z_k\right]$, we can simply choose a UAV position $\mathbf{u} = \left[x_k,y_k,h_u\right]$ right above $\mathbf{w}_k$, where $h_u$ is sufficiently large for UAV deployment, e.g., $h_u=300$. Then, we compute $det(H)$ at the given location $\mathbf{u}$. If $det(H)>0$,  we replace (\ref{con:accuracy}) in (P1) with (\ref{con:temp2}) for user $k$, otherwise we replace (\ref{con:accuracy}) with (\ref{con:temp3}). For simplicity of notation, we denote the localization-feasible region of the UAV as
	\begin{equation}
		\mathbf{u}\in\mathcal{A}_k(\hat{P}), \forall k\in\{1,2,\cdots,K\}, \label{region}
	\end{equation}
	where $\mathcal{A}_k(\hat{P})$ is the 3D feasible conic region within which the UAV satisfies the positioning accuracy requirements of user $k$ under a given localization power $\hat{P}=\{\bar{{P}_{1}},\bar{{P}_{2}},\bar{{P}_{3}}\}$ of the BSs.
	
	\begin{figure}[t]
		\centering
		\begin{minipage}[t]{0.24\linewidth}
			\centering
			\includegraphics[width=\textwidth]{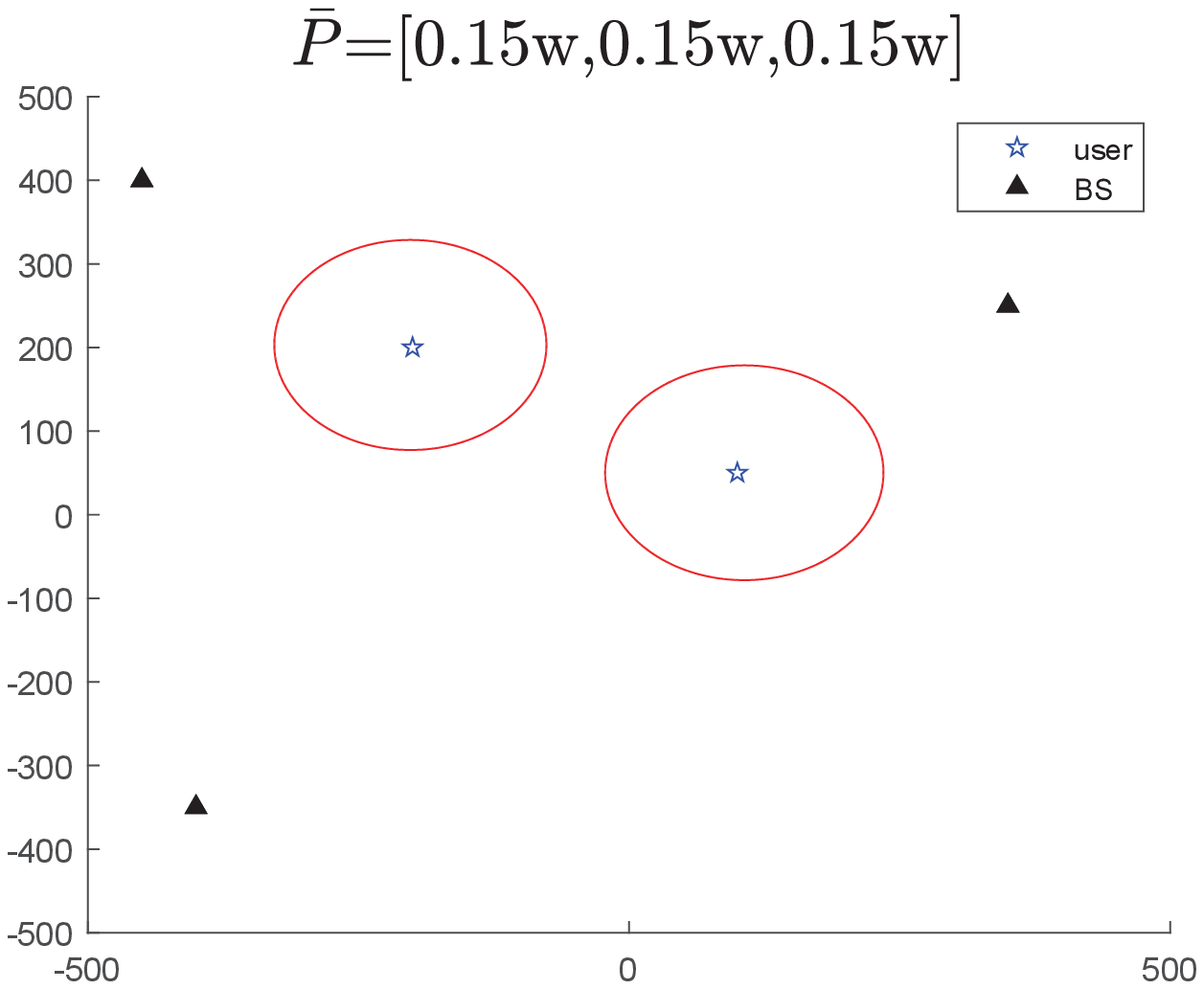}
		\end{minipage}%
		\begin{minipage}[t]{0.24\linewidth}
			\centering
			\includegraphics[width=\textwidth]{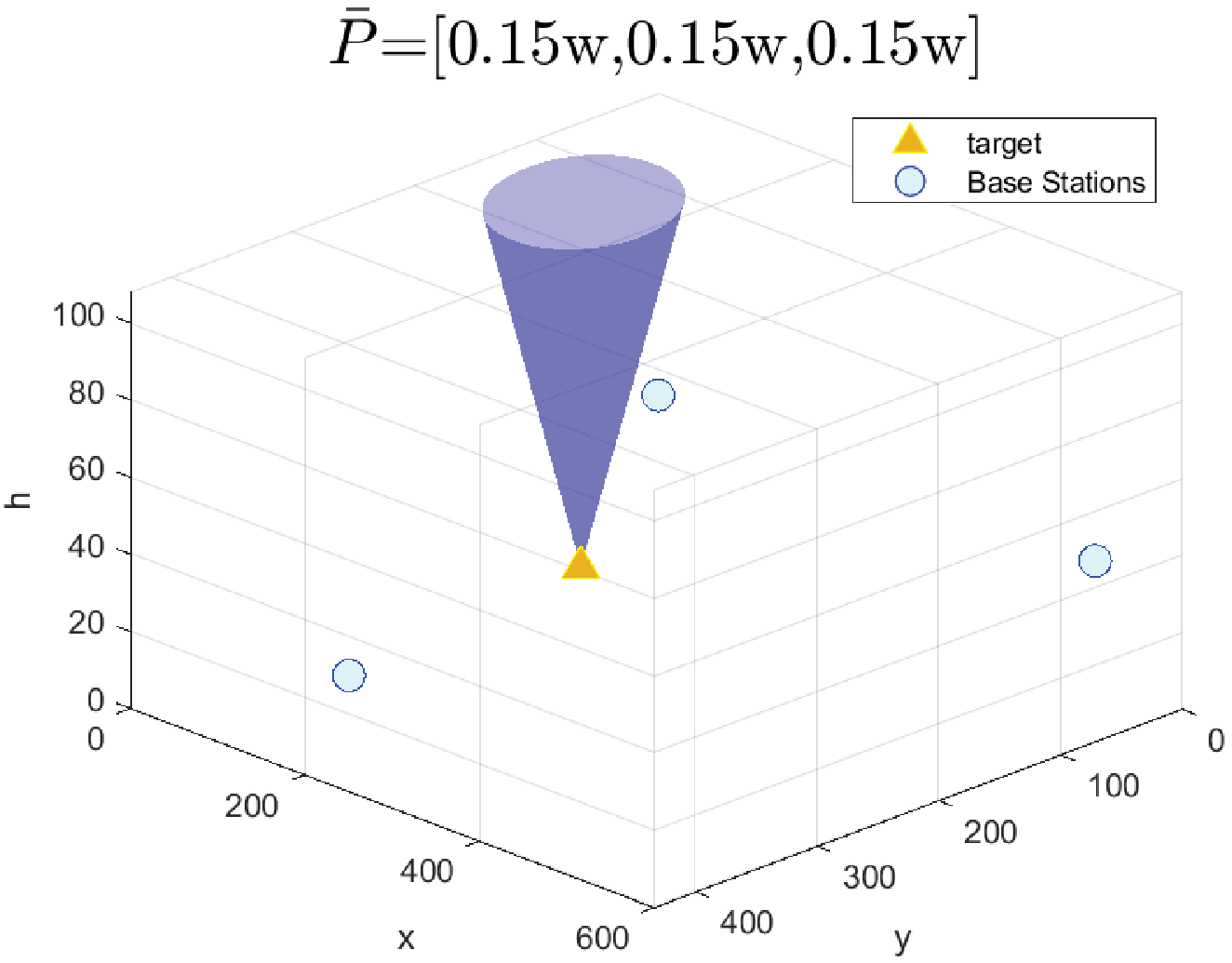}
		\end{minipage}%
		\begin{minipage}[t]{0.24\linewidth}
			\centering
			\includegraphics[width=\textwidth]{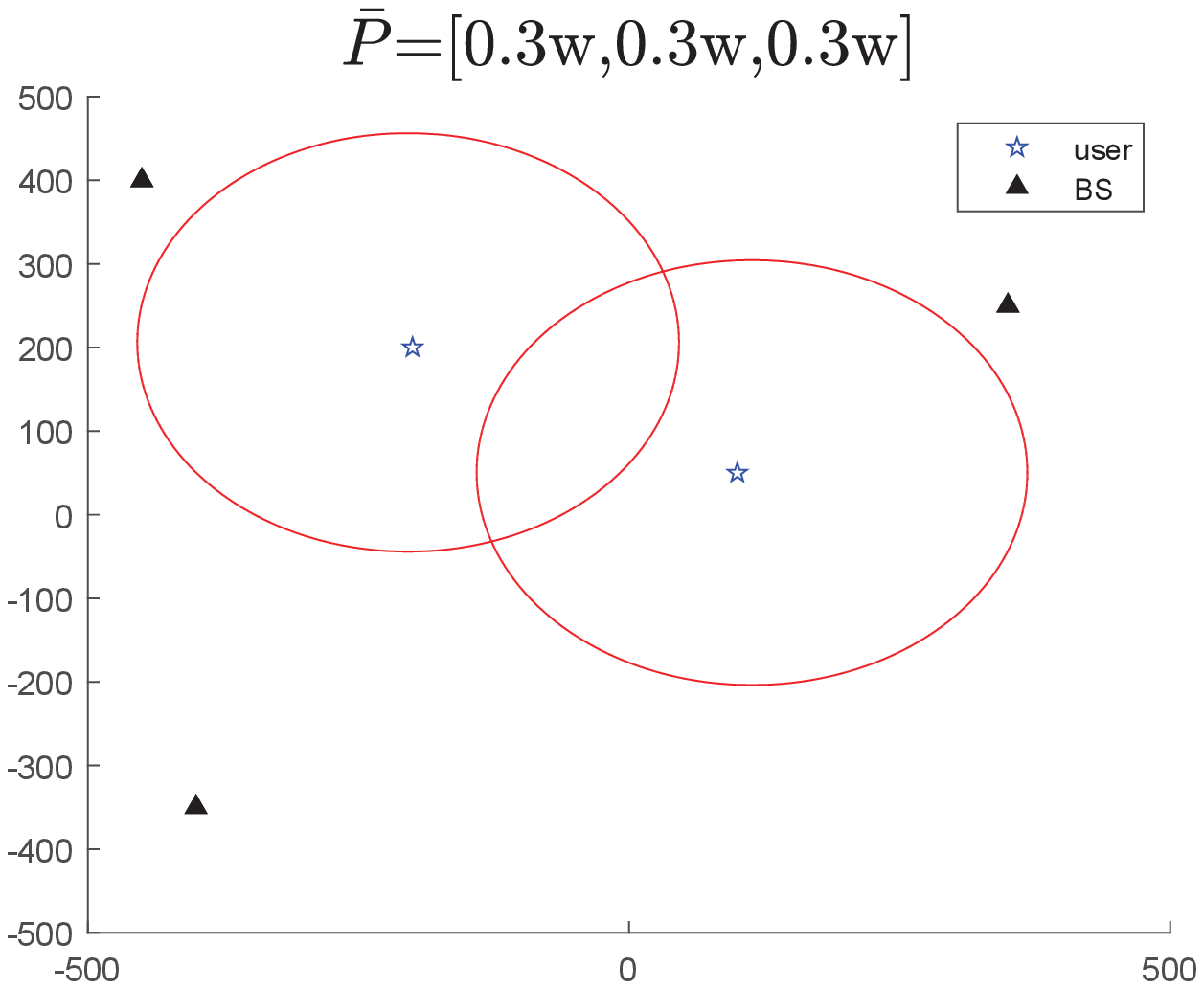}
		\end{minipage}%
		\begin{minipage}[t]{0.24\linewidth}
			\centering
			\includegraphics[width=\textwidth]{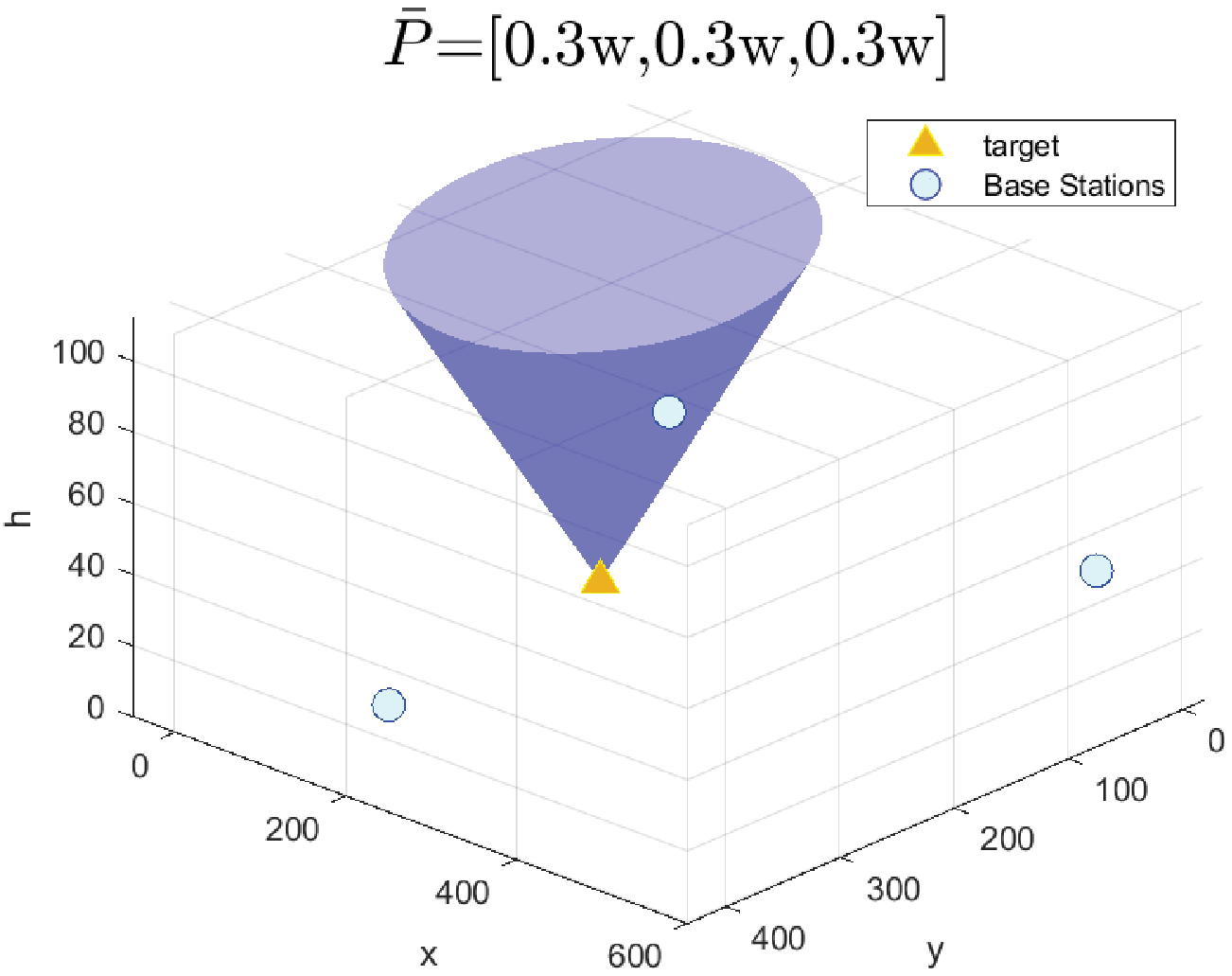}		
		\end{minipage}
		\caption{The 2D feasible ellipse regions of UAV flying at 200m and 3D feasible conic regions under different positioning power $\hat{P}$ of BSs. }
		\label{fig33}
	\end{figure}
	
	Fig. \ref{fig33} shows the feasible area of UAV under different BS positioning power. The 2D feasible region becomes larger as the positioning power increases or the altitude of the UAV increases. Thus, we need to jointly optimize the resource allocation and UAV 3D position to attain an optimum performance.
	\section{Proposed Algorithm}
	After replacing (\ref{con:accuracy}) with (\ref{region}),  Problem (P1) becomes:
	\begin{subequations}
		\begin{align}
			\mathrm{(P2):}
			\quad &  \underset{{P},{S},\mathbf{u},{\hat{P}}}{\text{max}} \ \
			  \sum_{k=1}^{K}  R_{k}  \nonumber \\
			s.t.\quad & (\ref{con:power}),(\ref{con:bandwidth}),(\ref{con:rate}),(\ref{con:fea}),(\ref{region}). \nonumber 
		\end{align}	
	\end{subequations}	
	Problem (P2) is still non-convex and we develop in the following an efficient algorithm based on Gibbs sampling, where the pseudo-code is summarized in Algorithm 1.

	\subsection{Bandwidth Allocation and Power Optimization}
	Given a feasible UAV's position $\mathbf{u}$ and positioning power ${\hat{{P}}}$, Problem (P2) reduces to:
	\begin{subequations}
		\begin{align}
			\mathrm{(P2\text{-}BAPO):\ }
			\quad &  \underset{{P},{S}}{\text{max}}\  \sum_{k=1}^{K} R_{k} \nonumber\\
			s.t.\quad & (\ref{con:power}),(\ref{con:bandwidth}),(\ref{con:rate}), (\ref{con:fea}),\nonumber
		\end{align}	
	\end{subequations}
	which is convex and can be solved by off-the-shelf convex algorithms. To gain insights into the optimal solution structure, we  derive partial Lagrangian of (P2-BAPO) as
	\begin{equation}
\label{2}
		\begin{aligned}
			&\mathcal{L}(P,S,\lambda,\mu,\nu)\\
			&= \sum_{k=1}^{K} R_{k} -  \sum_{j\in\{1,2,3,u\}}\lambda_j \left( \sum_{k=1}^{K}P_{jk}+\bar{P_{j}}-P_{max} \right) \\
			& - \sum_{j\in\{1,2,3,u\}} \mu_j \left( \sum_{k=1}^{K}s_{jk}-1 \right)- \sum_{k=1}^{K}\nu_k \left(R_{th}-R_{k} \right),
		\end{aligned}
	\end{equation}
	where $\mathbf{\lambda}=[\lambda_1,\lambda_2,\lambda_3,\lambda_u]$, $\mathbf{\mu}=[\mu_1,\mu_2,\mu_3,\mu_u]$, $\mathbf{\nu}=[\nu_1,\nu_2,\cdots,\nu_K]$ are the Lagrangian multipliers to the constraints. The dual function can be expressed as
	\begin{equation}
		\begin{aligned}
			d(\mathbf{\lambda},\mathbf{\mu},\mathbf{\nu}) = \underset{P,S}{\text{max}} \ \mathcal{L}\  \   s.t.\ \  P_{jk},s_{jk} \geq 0, \forall j,k,  \label{dualfun}
		\end{aligned}	
	\end{equation}
	and the dual problem is $ \text{min} \left\{d(\mathbf{\lambda},\mathbf{\mu},\mathbf{\nu})| \mathbf{\lambda},\mathbf{\mu},\mathbf{\nu}\geq \mathbf{0} \right\}.$ Given a set of dual variables, we take the derivative of $\mathcal{L}$ with respect to $s_{jk}$ and set to zero, the optimal solution $\{P^*,S^*\}$ satisfies the following first-order KKT condition \cite{Bi2020}
\begin{equation}
\frac{s_{jk}}{P_{jk}} = \frac{h_{jk}}{ -\left( W\left(-\frac{1}{ \exp \left(1 + \frac{\mu_j \ln 2}{B\left(1+\nu_k\right)}\right)}\right)\right)^{-1}-1}, \forall j,k,  \label{con:PS}
\end{equation}
where $h_{jk} = \frac{F^{-1}\left(\epsilon;2,\lambda_G\right) \beta}{B N_0 d_{jk}^\iota}$ for the G2G channel, and $h_{uk} = \frac{F^{-1}\left(\epsilon;2,\lambda_A\right) \beta}{B N_0 d_{uk}^{\tilde{\iota}}}$. $W(x)$ denotes the Lambert-W function, which is the inverse function of $f(y)=ye^y=x$, i.e., $y=W(x)$.
	
	Notice that any solution $\{P^{'},S{'}\}$ satisfies (\ref{con:PS}) is an optimal solution to (\ref{dualfun}), thus there are an infinite number of equally optimal solutions. We can easily find a solution $\{P^{'},S^{'}\}$ that satisfies the equalities constraints (\ref{con:power}), (\ref{con:bandwidth}) and (\ref{con:PS}) by solving an under-determined system of linear equations, which contains $(K+2) \times 4$ equations and $8K$ variables $(K\geq 2)$. By substituting $\{P^{'}, S{'}\}$ into (\ref{rateG}) and (\ref{rateA}), we can calculate the data rate $R_k$ for each user, with which we can obtain the sub-gradients of the dual variables $\{\mathbf{\lambda},\mathbf{\mu},\mathbf{\nu}\}$ from (\ref{2}). Here, the sub-gradients of $\mathbf{\lambda},\mathbf{\mu}$ are always $0$ because of the linear equality constraints (\ref{con:power}) and (\ref{con:bandwidth}) enforced. Therefore, we only need to iteratively update $\mathbf{\nu}$ by the sub-gradient method until a convergence condition is met, e.g.,	the change of $||\mathbf{\nu}||$ is marginal. After obtaining the optimal $\mathbf{\nu ^{*}}$, we substitute (\ref{con:PS}) into (P2-BAPO)
	and reduce the problem to  simple linear programming, which can be solved  by the simplex method and lead to an optimal solution $\{P^*, S^*\}$.
	\begin{algorithm}[t]
		\caption{The proposed method for solving (P2). }
		\KwIn{ positioning accuracy $\mathbf{\epsilon}$, user target position $\mathbf{w}$;}
		\KwOut{UAV's position $\mathbf{u}$, communication power ${P}$, bandwidth allocation ${S}$,  positioning power of BSs $\hat{P}$; }
		\textbf{Initialize}
		{ $ {\hat{P}}=\{2\Delta P,2\Delta P,2\Delta P\}$, parameter $T$ and $\alpha <1$;}\\
		\Repeat
		{
			the objective value converges
		}
		{
			Create candidate positioning power sets $\hat{P}^{'}$ based on $\hat{P}$;\\
			\For {any ${\hat{P}^{j}} \in \hat{P}^{'}$}
			{	Obtain the localization-feasible region of UAV in (\ref{region}) given $\hat{P}^j$; \\
				\textbf{Initialize}
				{A feasible UAV position $\mathbf{u}$ that satisfies (\ref{region});}\\
				
				\Repeat
				{improvement of objective is sufficiently small}
				{	Obtain optimal  $\{P, S\}$ by solving (P2-BAPO);\\
					Update $\mathbf{u}$ by solving (P2-UDO) given $\{P,S\}$;\\
			}}
			If (P2) is infeasible for all $\hat{P}^j \in \hat{P}^{'}$, update $\hat{P}$ with (\ref{trans});\\
			Otherwise, update $\hat{P}$ based on the transfer probability defined by (\ref{probability});\\
			$T \leftarrow \alpha T$;}
	\end{algorithm}

	\subsection{UAV Deployment Optimization}
	Given $\{\hat{P}, P, S\}$, the UAV's position can be  updated by solving the  following problem:
	\begin{subequations}
		\begin{align}
			\mathrm{(P2\text{-UDO}):\ }
			\quad &  \underset{\mathbf{u}}{\text{max}} \ \sum_{k=1}^{K} R_{k} \nonumber \\
			s.t.\quad & (\ref{con:rate}),(\ref{region}) \nonumber
		\end{align}	
	\end{subequations}
	Although  (P2-UDO) is non-convex, it can be solved efficiently by the successive convex approximation (SCA) technique that repeatedly linearizes the rate expression $R_k$ with respect to $\mathbf{u}$ \cite{ref17}. We omit the detail of SCA due to the page limit. Thus, for any given positioning power we can optimize $\{P,S\}$ and $\{\mathbf{u}\}$ by iteratively solving (P2-BAPO) and (P2-UDO), until the improvement of objective value is less than a threshold.

	\subsection{Positioning Power Optimization}
	To tackle the non-convex optimization of positioning power $\hat{P}$, we propose a Gibbs sampling (GS) based method that iteratively updates $\hat{P}$ in a probabilistic manner. The GS method generates a series of positioning power samples  and iteratively updates  the  power  according to a transfer probability. Firstly, we discretize the value of positioning power as $\{0,\Delta P,2\Delta P,\cdots,  P_{max}\}$, where $\Delta P$ is the step size. Because constraint (\ref{region}) is unrelated to UAV's positioning power, we set a fixed $\bar{P}_u$ such that the opt$-$D approximation in Fig. 1 is accurate. Let $\hat{P}=\{\bar{{P}_{1}},\bar{{P}_{2}},\bar{{P}_{3}} \}$ denote the current positioning power of the three BSs. $J(\hat{P})$ denotes the optimal objective value of (P2) given $\hat{P}$ and $\bar{P}_u$. If $(P2)$ is infeasible, $J(\hat{P})=-\infty$. Then, we generate a candidate positioning power set $\hat{P}^{'}$ by  varying at most one component of $\hat{P}$ at a time, for example, we alter $\bar{P}_1$ and get a new positioning power $\{\bar{{P}_{1}} +\Delta P,\bar{{P}_{2}},\bar{{P}_{3}} \}$ or $\{\bar{{P}_{1}} -\Delta P,\bar{{P}_{2}},\bar{{P}_{3}} \}$. Therefore, we get a set $\bar{P}^{'}$ containing the origin positioning power $\hat{P}$ and its 6 neighboring positioning power vector. For $\hat{P}^{j}\in \hat{P}^{'}$, its transfer probability can be expressed as
	\begin{equation}
		f(\hat{P}^{j})=\frac{\exp(J(\hat{P}^{j})/T)}{\sum_{\hat{P}^{i}\in\hat{P}^{'}}\exp(J(\hat{P}^{i})/T)}, \label{probability}
	\end{equation}
	where  $T>0$ is a fixed temperature parameter.  If the problem is infeasible for all $\hat{P}^j \in \hat{P}^{'}$, we increase the  positioning power:
	\begin{equation}
		\hat{P}=\text{min} \left\{ (\bar{{P}_{1}}+\Delta P,\bar{{P}_{2}}+\Delta P,\bar{{P}_{3}}+\Delta P ), P_{max}\right\}. \label{trans}
	\end{equation}
	From (\ref{probability}), the positioning power tends to converge to $\bar{P}^{j}$ that yields a higher objective value, especially when $T$ is small. The stochastic transfer process can avoid falling into a local optimum prematurely. In addition, to speed up the convergence, we update $T$ in each iteration to $\alpha T$, where $\alpha<1$.

\section{Simulation Results}
	In this section, we evaluate the performance of the proposed method by simulations. All the computations are solved in Pycharm on a computer with an AMD R7-5800H 3.20-GHz CPU and 16 GB of memory. We consider a 1 km $\times$ 1 km square area, where the coordinates of the BSs are (-400,-350,10), (-450,400,10) and (350,250,10). The coordinates of $7$ target users are (-60,-110,12), (150,-70,29), (-350,30,22), (-140,-60,26), (-250,130,15), (-280,-210,17) and (-220,260,32). The simulation parameters are $B=1$MHz, $\bar{B}=180$KHz, $N_0= -157$dbm/Hz, $\beta= -38.89$dB, $R_{th} = 2.5$Mbps, $\sigma_{nlos}^{2}= 6 \times 10^{-18} s^2$, $\psi = 5.8\times 10^{-16}s^2$, $T = 0.95$, $\iota= 2.3$ and $\tilde{\iota} = 2$. $\omega_G = 1$ and $\omega_A = 0.2$, $\varepsilon=0.1$, such that $F^{-1}(\cdot) = 0.11$ and $0.32$ for the G2G and A2G channels, respectively. Let $lb_k$ and $ub_k$ denote the lower and upper bounds of feasible $\epsilon_k$, specified either in (\ref{con:fea1}) or (\ref{con:fea2}) by setting $\bar{P_i} = 0.15$ w. We set $\epsilon_k = lb_k + \zeta(ub_k-lb_k)$, where $\zeta\in (0,1)$ is a tunable parameter. Unless otherwise stated, we set $P_{max} =1$w, $\zeta = 0.7$, and $\bar{P}_u = 0.2$ w such that $opt$-$D_1 \approx$ $opt$-$D$ holds in the considered setup.

We consider the following benchmark methods: 1) Particle Swarm Optimization (PSO)\cite{ref9}: a meta-heuristic algorithm that iteratively searches the vector space, which approaches the global optimum with large-scale implementation. 2) Equal Power Allocation (EPA): each BS or UAV allocates half of the available power to positioning service, and allocates the remaining power equally to all the users for communications. 3) UAV Center Deployment (UCD)\cite{ref17}: UAV is deployed at the geometric center of the 2D positions of the target users with an altitude of $500$m. For EPA and UCD methods, the other variables are optimized using the method in Section IV.

\begin{figure}[t]
		\centering
		\begin{minipage}[t]{0.22\linewidth}
			\centering
			\includegraphics[width=\textwidth]{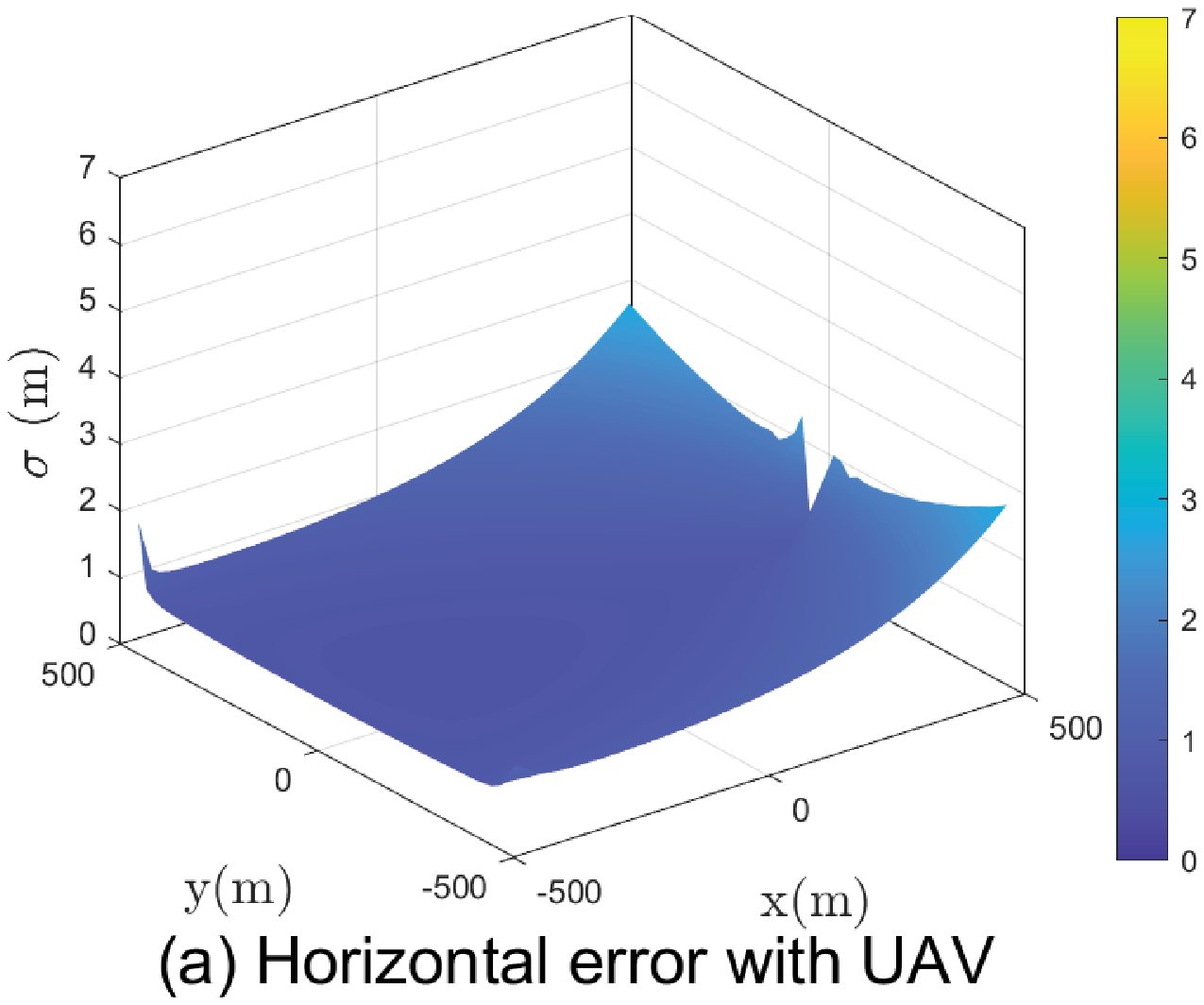}
		\end{minipage}%
		\begin{minipage}[t]{0.24\linewidth}
			\centering
			\includegraphics[width=\textwidth]{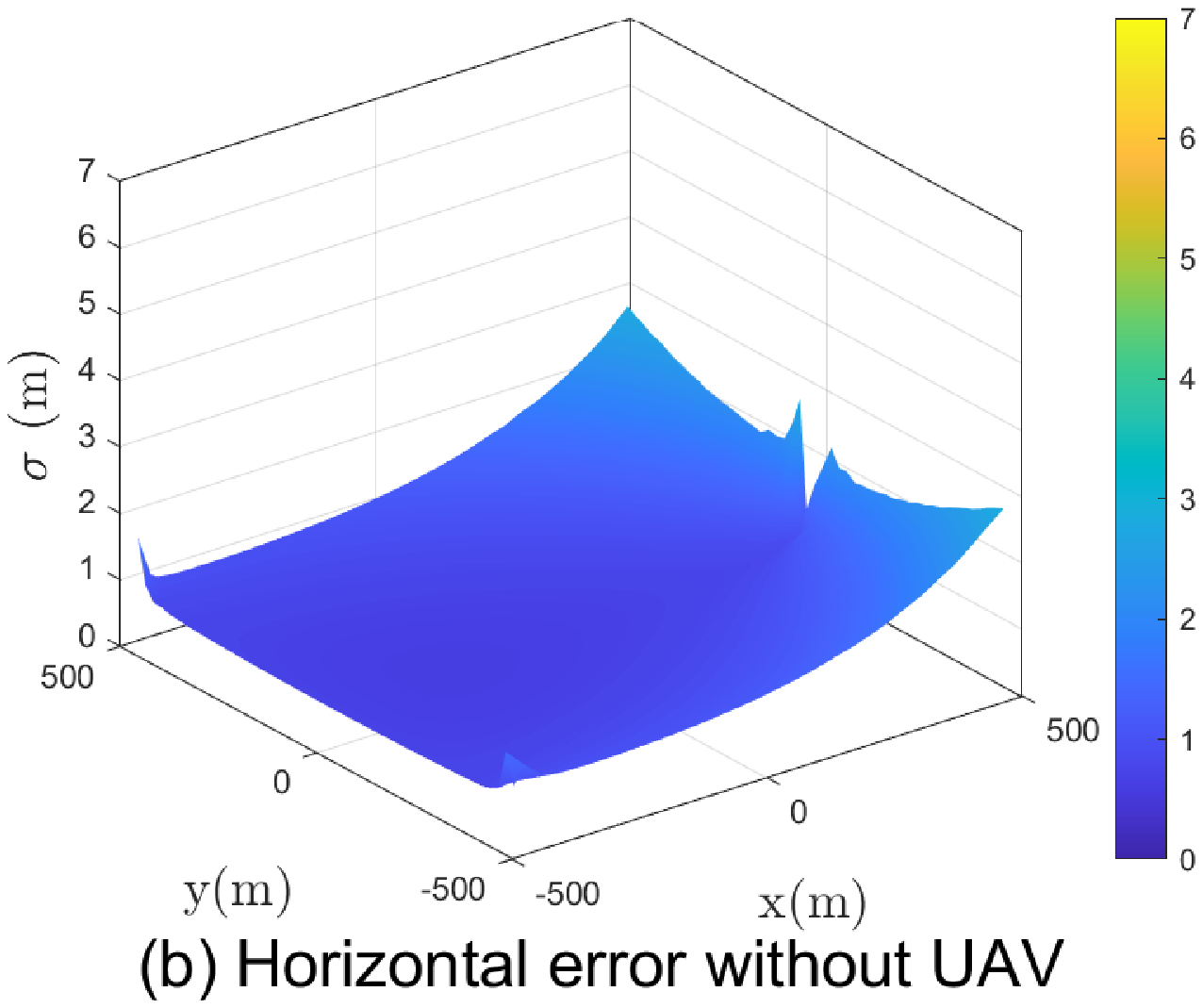}
		\end{minipage}%
		\begin{minipage}[t]{0.24\linewidth}
			\centering
			\includegraphics[width=\textwidth]{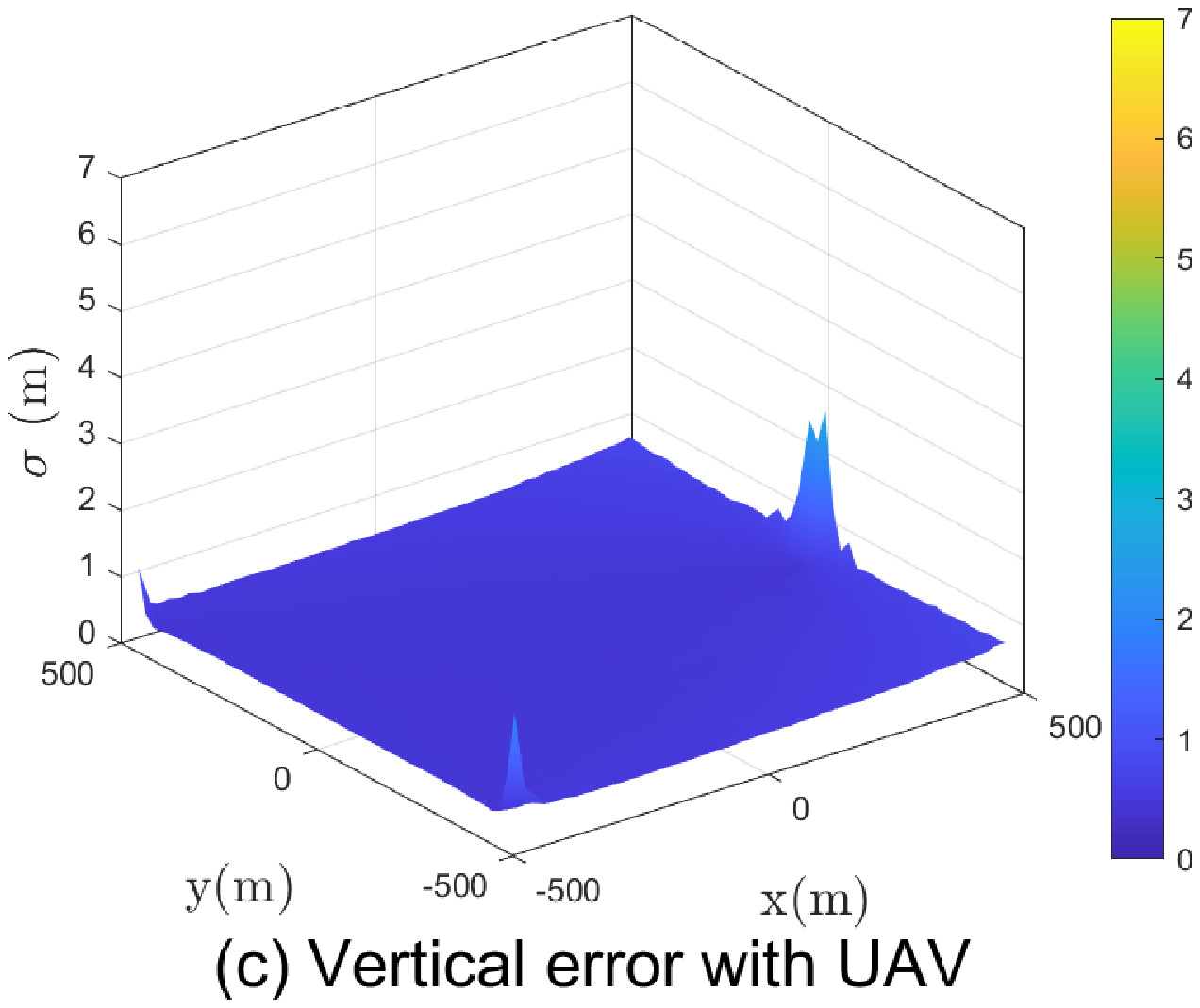}
		\end{minipage}%
		\begin{minipage}[t]{0.24\linewidth}
			\centering
			\includegraphics[width=\textwidth]{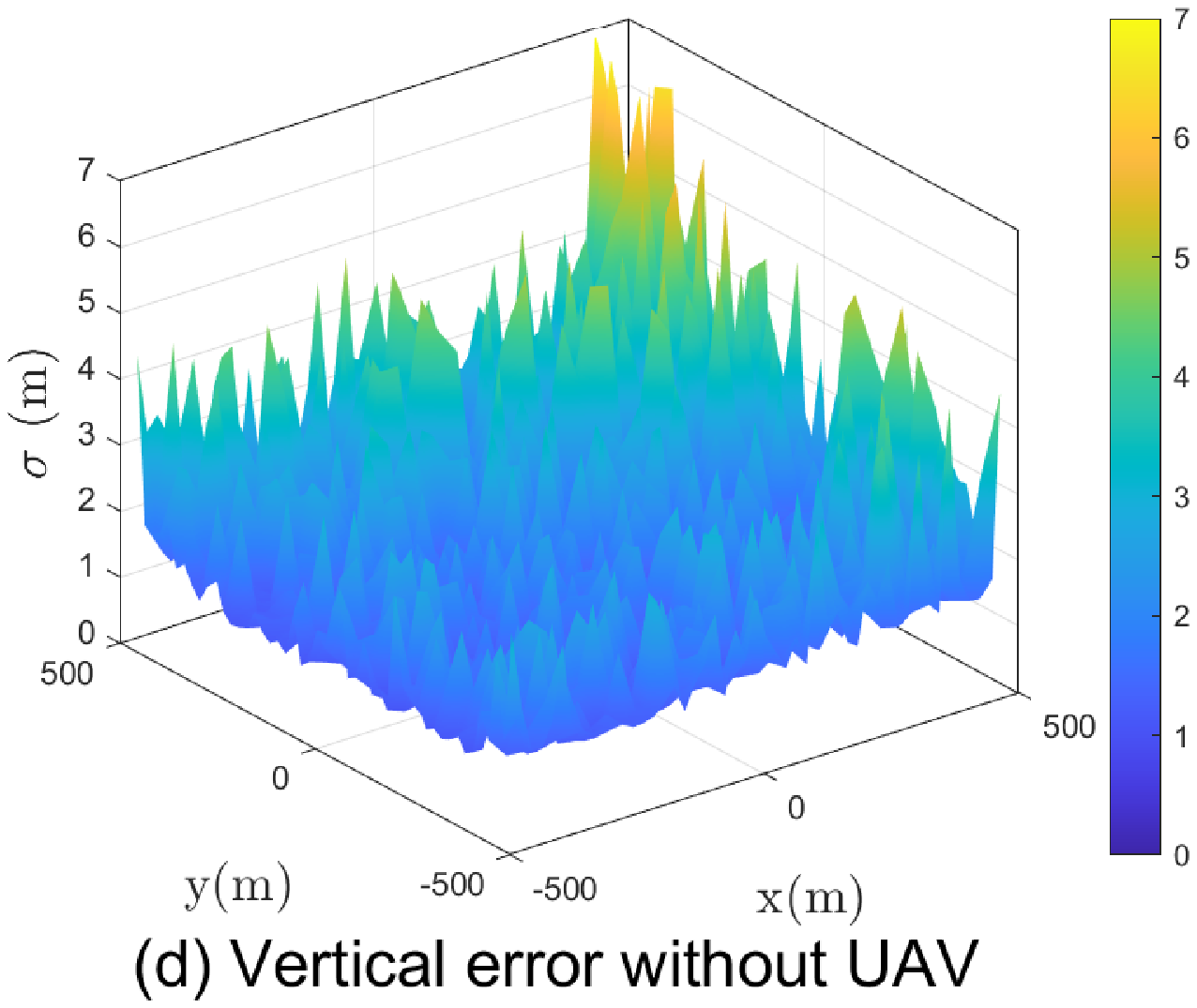}		
		\end{minipage}
		\caption{Comparisons of positioning error (CRLB) with and without the assistance of UAV in horizontal and vertical directions. }
		\label{fig5}
	\end{figure}

In Fig.~\ref{fig5}, we highlight the performance advantage of UAV-assisted localization scheme compared with the conventional ground anchor scheme, especially in the vertical direction. In particular, we fix the positions of three ground anchors and optimize the position of the fourth anchor to minimize the 3D localization error at ground target locations. The UAV-assisted scheme and the conventional scheme employ a UAV anchor (at fixed 100 meters altitude) and a ground anchor (at fixed 10 meters altitude) as the fourth anchor, respectively. The positioning power of all anchors is 1w. We divide the network into 10m $\times$ 10m squares, and for each grid point, we exhaustively search the optimal horizontal position of the fourth anchor that minimizes the CRLB of the 3D positioning error. In Fig.~\ref{fig5}(a) and (b), we compare the horizontal localization errors and find that both schemes produce similar errors within $[0.6,2.8]$ meters. This shows that the horizontal positioning performance is insensitive to altitude of the fourth anchor. However, the UAV-assisted method reduces vertical positioning error of the conventional scheme from $[1.1, 7.1]$ meters to $[0.4, 2.7]$ meters, thanks to the amplitude diversity provided by the UAV anchor. This shows that UAV-assisted positioning networks can substantially improve the accuracy in the vertical direction without degrading the horizontal positioning accuracy, resulting in an overall improved 3D positioning performance.

\begin{figure}[t]
		\centering
		\includegraphics[width=0.6 \textwidth]{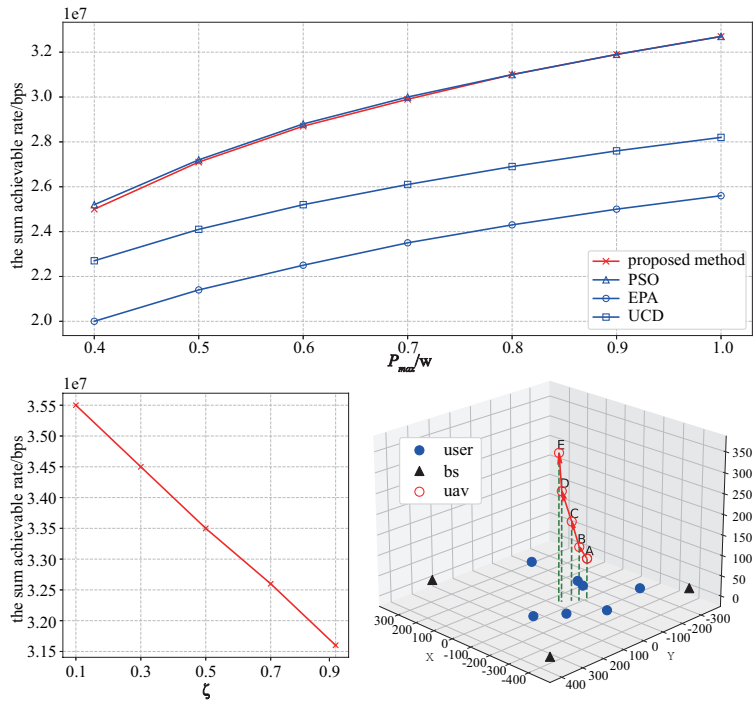}
		\caption{Performance comparisons and optimal UAV deployment. The UAV positions A to E in the bottom right figure correspond to $\zeta=\{0.1,0.3,0.5,0.7,0.9\}$, respectively. }
		\label{fig4}
	\end{figure}

In Fig.~\ref{fig4}, we then compare the performance of the proposed and benchmark methods when the maximum power $P_{max}$ varies. We see that the proposed method is almost identical to that of the PSO method when $P_{max}$ is larger than $0.4$w. When $P_{max}=$ 1w, the proposed method outperforms EPA and UCD methods by $15.9\%$ and $27.7\%$, respectively. This demonstrates the advantage of the proposed joint UAV placement and resource allocation optimization method. We then vary the localization precision requirement by tuning $\zeta$, where a larger $\zeta$ indicates higher precision requirement $\epsilon_k$. In the bottom left figure, the sum rate performance decreases with $\zeta$ as expected. In the bottom right figure, we also plot the optimal UAV positions under different $\zeta$. We see that, under a small $\epsilon_{k}$, the UAV tends to fly at a lower altitude to maximize the communication rate, however, as we increase $\epsilon_{k}$, the UAV tends to fly at a higher altitude to provide anchor location diversity for meeting higher vertical localization accuracy requirement. Then, we examine the algorithm complexity by considering networks consisting of $m$ randomly selected users from the $7$-user network. When $m$ increases from $2$ to $7$, we find that the number of search iterations consumed by the proposed GS-based method is almost constant, while that increases quickly from $102$ to $807$ by the PSO benchmark method. This is because the proposed GS method only optimizes the positioning powers of the three ground BSs, which does not scale with the user number. In terms of CPU time, the GS-based method increases mildly from $35.2$ to $47.0$ seconds, while that of the PSO method increases quickly from $62.5$ to $440.7$ seconds. Specifically, for the $7$-user case, the proposed method reduces the CPU time of the PSO benchmark by $89.3\%$. Overall, the proposed method is a more suitable solution under time-critical emergent situations.

	\section{Conclusions}
	In this paper, we studied the joint optimization of 3D deployment and resource allocation in a UAV-assisted integrated communication and localization network. Thanks to the assistance of the UAV,   positioning accuracy can be effectively improved. We first derived a closed-form expression of the localization accuracy constraint based on a new localization performance metric and characterized a convex geometric feasible region of UAV 3D deployment. Accordingly, we proposed an efficient method to solve the non-convex joint deployment and resource allocation problem. Numerical results showed that the proposed method attains almost identical rate performance as the meta heuristic benchmark method while reducing the CPU time by $89.3\%$.

\end{document}